# Some Cosmological Consequences of the Correlation between the Gravitational and Inertial Mass

Fran De Aquino      physics/0205040

Maranhao State University,
Physics Department,
65058-970 S.Luis/MA, Brazil.
E-mail: deaquino@elo.com.br

## Abstract

It has been demonstrated (gr-qc/9910036) that the gravitational and inertial masses are correlated by a *dimensionless electromagnetic factor*, which can be different from one. It can be also reduced, nullified or made negative in specific electromagnetic conditions. This unexpected theoretical result has been confirmed by an experiment using Extra-Low Frequency(ELF) radiation on ferromagnetic material (gr-qc/0005107). Recently another experiment using UV light on phosphorescent plastic has confirmed the phenomenon. Here we show some fundamental consequences of the correlation for the Quantum Cosmology.

## Introduction

In a previous paper[1] we have shown that the *gravitational mass* and the *inertial mass* are correlated by a dimensionless electromagnetic factor. The general expression of correlation between gravitational mass $m_g$ and inertial mass $m_i$, can be written as follows

$$\frac{m_g}{m_i} = 1 - 2\left\{\sqrt{1+\left\{\frac{q}{m_i c}\right\}^2} - 1\right\} \quad (1)$$

The *momentum* $q$ in the equation above, is given by

$$q = N\hbar k = N\hbar\omega/(\omega/k) = U/(dz/dt) =$$
$$= U/v \quad (2)$$

where $U$ is the electromagnetic energy absorbed (or emitted) by the particle; $v$ is the velocity of the incident( or emitted) radiation. It can be shown that

$$v = \frac{c}{\sqrt{\frac{\varepsilon_r \mu_r}{2}\left(\sqrt{1+(\sigma/\omega\varepsilon)^2}+1\right)}} \quad (3)$$

where ; $\varepsilon$, $\mu$ and $\sigma$, are the electromagnetic characteristics of the outside medium around the particle in which the incident radiation is propagating ( $\varepsilon = \varepsilon_r \varepsilon_0$ where $\varepsilon_r$ is the *relative electric permittivity* and $\varepsilon_0 = 8.854 \times 10^{-12} F/m$ ; $\mu = \mu_r \mu_0$ where $\mu_r$ is the *relative magnetic permeability* and $\mu_0 = 4\pi \times 10^{-7} H/m$ ). For an *atom* inside a body , the incident(or emitted) radiation on this atom will be propagating inside the body , and consequently , $\sigma = \sigma_{body}$ , $\varepsilon = \varepsilon_{body}$, $\mu = \mu_{body}$.

By substitution of Eqs.(2) and (3) into Eq.(1), we obtain

$$\frac{m_g}{m_i} = 1 - 2\left\{\sqrt{1 + \left\{\frac{U}{m_i c^2}\sqrt{\frac{\varepsilon_r \mu_r}{2}\left(\sqrt{1+(\sigma/\omega\varepsilon)^2}+1\right)}\right\}^2} - 1\right\}$$

$$= 1 - 2\left\{\sqrt{1 + \left\{\frac{U}{m_i c^2}n_r\right\}^2} - 1\right\} \quad (4)$$

In the equation above $n_r$ is the *refractive index* ($n_r = c/v$).

In this work we will show that the correlation above (Eq.4) has fundamental consequences for the Cosmology.

# 1. Cosmology

Initially let us now consider the problem of the *anomalies* in the spectral red-shift of certain galaxies and stars.

Several observers have noticed red-shift values that cannot be explained by the Doppler-Fizeau effect or by the Einstein effect (the gravitational spectrum shift, supplied by Einstein's theory).

This is the case of the called *Stefan's quintet* (a set of five galaxies which has been discovered in 1877), whose galaxies are located at approximately the same distance from the Earth, according to very reliable and precise measuring methods. But, when the velocities of the galaxies are measured by its red-shifts, the velocity of one of them is much greater than the velocity of the others.

Similar observations have been made on the *Virgo constellation* and spiral galaxies.

Also the Sun presents a red-shift greater than the predicted value by the Einstein effect.

We believe that all these anomalies can be explained if we consider the Eq.(4), for the calculation of the *gravitational mass* of the point of emission, before to use the well-known expression below of the gravitational spectrum shift, supplied by Einstein's theory [2]

$$\Delta\omega = \omega_1 - \omega_2 = \frac{\phi_2 - \phi_1}{c^2}\omega_1 =$$

$$= \frac{-Gm_{g2}/r_2 + Gm_{g1}/r_1}{c^2}\omega_1 \quad (5)$$

where $\omega_1$ is the frequency of the light at the point of emission ; $\omega_2$ is the frequency at the point of observation; $\phi_1$ and $\phi_2$ are respectively, the Newtonian gravitational potentials at the point of emission and at the point of observation.

The Eq. (5) has been deduced from the expression [3] $t = t_0\sqrt{-g_{00}}$ between *own time* (real time), $t$, and the temporal coordinate $x^0$ of the space-time ($t_0 = x^0/c$).

When the gravitational field is *weak*, the temporal component $g_{00}$ of the metric tensor is given by [4] $g_{oo} = -1 - 2\phi/c^2$. Thus we can write

$$t = t_0\sqrt{1 - 2Gm_g/rc^2} \quad (6)$$

Curiously, the Eq. (6) tell us that we can have $t < t_0$ when $m_g > 0$ ; $t = t_0$ if $m_g = 0$ and $t > t_0$ for $m_g < 0$. In addition, if $m_g = c^2 r/2G$, i.e., if $r = 2Gm_g/c^2$ ( *Schwarzschild radius*[5] ) we obtain $t = 0$.

These theoretical results lead to a new *time* conception with amazing practical consequences.

Let us now consider the well-known process of stars' *gravitational contraction*. It is known that the destination of the star is directly correlated to its mass. If the star's mass is less than $1.4M_\odot$ ( Schemberg-Chandrasekhar's limit), its becomes a *white dwarf*. If its mass exceeds that limit, the pressure produced by the degenerate state of the matter no longer counterbalances the gravitational pressure, and the star's contraction continues. Afterwards occurs the reactions between protons and electrons ( capture of electrons ), where *neutrons* and anti-neutrinos are produced.

The contraction continues until the system regains stability ( when the pressure produced by the neutrons is sufficient to stop the gravitational collapse). Such systems are called *neutron stars*. There is also a critical mass for the stable configuration of neutron stars. This limit has not been fully defined as yet, but it is known that it is located between $1.8M_\odot$ and $2.4M_\odot$. Thus, if the mass of the star exceeds $2.4M_\odot$, the contraction will continue.

Note that at the end of the gravitational contraction of stars, there is a natural convergence towards a *neutrons cluster* in which the neutrons are continuously compressed.

We know that the neutrons have a *spin magnetic field* $\vec{H}_n$ which we write in the form

$$\vec{H}_n = \frac{\vec{M}_n}{2\pi r^3} \quad (7)$$

where

$$\vec{M}_n = \gamma_n \left(\frac{e}{2m_n}\right)\vec{S}_n$$

is the *spin magnetic momentum*; $\vec{S}_n = \hbar\sqrt{s(s+1)} = \frac{\sqrt{3}}{2}\hbar$ and $\gamma_n = 3.8256$ is the *gyromagnetic ratio* for the neutrons[6].

Therefore the Eq.(7) can be written as follows

$$\vec{H}_n = \gamma_n \left(\frac{\sqrt{3}}{8\pi}\right)\left(\frac{e\hbar}{m_n r^3}\right) \quad (8)$$

Thus, each neutron of neutrons cluster will be spinning in the magnetic field produced by the others neutrons of the cluster.

It is known that particles with electric charge $Q$ spinning in a magnetic field $H$ emit radiation with power $P$ which, as we know, is given by[7]

$$P = \frac{\mu^3 Q^4 H^2 V^2}{6\pi m_i^2 c(1 - V^2/c^2)} \quad (9)$$

Where $V$ is the tangential velocity of the particle.

Most of the emitted radiation has frequency $f$

$$f = f_0\left(1 - \frac{V^2}{c^2}\right)^{-3/2} \quad (10)$$

where

$$f_0 = \frac{\mu Q H}{2\pi m_i}\sqrt{1 - V^2/c^2} \quad (11)$$

is the named *cyclotron frequency*.

The neutrons do not have electric charge, but their *quarks* have. Consequently, they will emit radiation when they are spinning in a magnetic field. Thus the total radiation produced by the quarks will be the radiation emitted from the neutron.

Inside the neutrons, the tangential velocity $V$ of the quarks is

$$V = \omega r_0 = \frac{S_n}{I_n}r_0 = \left(\frac{\frac{\sqrt{3}}{2}\hbar}{\frac{2}{5}m_n r_n^2}\right)r_0 \quad (12)$$

where $S_n = I_n\omega = \frac{\sqrt{3}}{2}\hbar$ is the spin angular momentum of the neutron and $I_n$ its momentum of inertia; $r_n = 1.4\times 10^{-15}m$ is the radius of the neutron; $r_0$ is the distance from the rotating axis to the *quark* ( see Fig. 1 ).

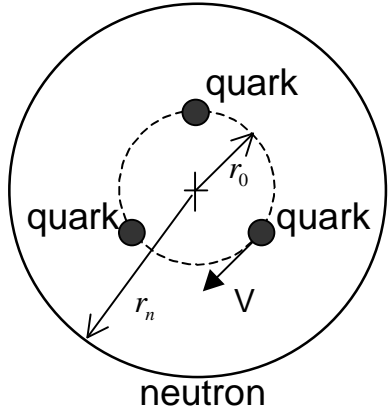

Fig. 1 - The neutron

Since $r_0 << r_n = 1.4 \times 10^{-15} m$ we can write from the Eq. (12) that $V << 6.97 \times 10^{22} r_n = 9,7 \times 10^7 m/s$ then we conclude that $V << c$.

In this way we can say that the radiation emitted from each *neutron* has power $P$ which, according to Eq. (9), is given by

$$P \cong \frac{\mu^3 e^4 H^2 V^2}{6\pi m_n^2 c} \qquad (13)$$

Then, according to Eq. (10), most of the emitted radiation from the neutrons has frequency

$$f = \frac{\mu e H}{2\pi m_n} \qquad (14)$$

The Eq.(13) shows that the power radiated from the neutron is
$$P << 10^{-32} H^2$$

Since we are not able to produce $H > 10^7 A/m$, we cannot observe the radiation emitted from the neutrons, simply because we are not able to detect radiation sources with $P << 10^{-18} W$. However, there is a way to check the radiation emitted from the neutrons. Consider a cluster of neutrons with $\sim 10^{12}$ neutrons within a magnetic field $H \approx 10^7 A/m, (B \approx 10T)$.

The power radiated from the cluster will be $\sim 10^{-6} W$ at 191.5MHz and can be experimentally detected.

Now we will calculate the gravitational mass of a neutron at the *neutrons cluster*.

To simplify the calculation we can consider only the contribution of the emitted radiation with frequency $f = \frac{\mu e H}{2\pi m_n}$. Then we can put $U = nhf$ in Eq.(4), where $n$ is the number of emitted photons from the neutron. Thus we can write

$$m_{gn} = m_n - 2\left\{ \sqrt{1 + \left[ \frac{nhf}{m_n c^2} n_r \right]^2} - 1 \right\} m_n \qquad (15)$$

But $P = nhf / \Delta t = nhf^2$, thus we can write $n = P/hf^2$. Substitution of $n$ into Eq.(15) gives

$$m_{gn} = m_n - 2\left\{ \sqrt{1 + \left[ \frac{P}{m_n c^2 f} n_r \right]^2} - 1 \right\} m_n \qquad (16)$$

By the substitution of Eqs. (13) and (14) and $n_r = 1$ into equation above we obtain

$$m_{gn} = m_n - 2\left\{ \sqrt{1 + \left[ \left( \frac{\mu^2 e^3 H V^2}{3 m_n^2 c^3} \right) \right]^2} - 1 \right\} m_n \qquad (17)$$

in which case $H$ is the total magnetic field in the center of mass of the neutron. To simplify the calculation we will just consider the contribution $H_n$ of an only neighboring neutron. Substitution of $H$ by $H_n$ from the Eq. (8) into Eq. (17), gives

$$m_{gn} = m_n - 2\left\{ \sqrt{1 + \left[ \gamma_n \left( \frac{\sqrt{3}}{8\pi} \right) \left( \frac{\mu^2 e^4 \hbar V^2}{3 m_n^3 c^3 r^3} \right) \right]^2} - 1 \right\} m_n \qquad (18)$$

where $r$ is the distance between the centers of mass of the two neutrons.

At the *supercompressing* state, $r = 2r_n^*$ ($r_n^*$ is the radius of the neutron at the *supercompressing state*, see Fig.2). The angular velocity $\omega$ is *constant* (due to the spin $s = 1/2$). Thus, from Eq.(12), we can write

$$V = \omega r_0^* = \tfrac{5\sqrt{3}}{4}\left(\frac{\hbar}{m_n r_n^2}\right) r_0^* \qquad (20)$$

where $r_0^*$ is the radius $r_0$ at the *supercompressing state*.

Substitution of Eq.(20) into Eq.(18) leads to the following equation

$$m_{gn} = m_n - 2\left\{\sqrt{1 + \left[4.584\times 10^{-23}\frac{(r_0^*)^2}{(r_n^*)^3}\right]^2} - 1\right\}m_n \qquad (21)$$

The lower critical value for $r_0^*$ is the neutrons' *Schwarzschild radius*, $r_{sn}$, for $m_{gn} = m_n$, i.e., $r_{sn} = 2Gm_n/c^2 = 2.475\times 10^{-54}\,m$ (because if $r_0^* < r_{sn}$, will occur a singularity).

Thus, substitution of $r_0^* = 2.475\times 10^{-54}\,m$ into Eq. (21) leads to the gravitational mass of the neutrons at the *final stage* of the cluster's gravitational contraction. i.e.,

$$m_{gn} = m_n - 2\left\{\sqrt{1 + \left[0.28\times 10^{-129}\frac{1}{(r_n^*)^3}\right]^2} - 1\right\}m_n \qquad (22)$$

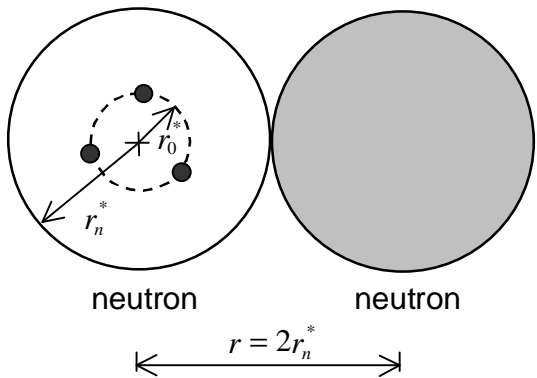

Fig.2 - Neutrons at the supercompressing state

From the equation above we conclude that the gravitational masses of the neutrons will be *negative* if $r_n^* < 6.3\times 10^{-44}\,m$.

From the *Quantum* Cosmology we know that: if the equation of state of the Universe is $p = \eta\rho$ (where $p$ is the pressure and $\rho$ is the density), then the length scale of the Universe $R$ increases with time $t$ according to [8]

$$R \propto t^{2/[3(1+\eta)]} \qquad (23)$$

In the Elementary Particle picture, $p = \tfrac{1}{3}\rho$, thus the most natural value of $\eta$ is $\tfrac{1}{3}$, then the size of the Universe before $10^{-35}\,s$ (Grand Unification era) was very less than the size of an elementary particle ($10^{-15}\,m$).

Let us imagine the Universe coming back for the past. There will be an instant in which it will be similar to a *neutrons cluster*, such as the stars at the final state of gravitational contraction. Consequently, assuming that at this stage the *average* density of the Universe was equal to the *average* density of the *supercompressing* neutrons we can write

$$\left(\frac{R}{\bar{r}_n^*}\right)^3 = \frac{M}{m_n} \qquad (24)$$

Where $M \approx 10^{53}\,kg$ is the inertial mass of the Universe. Thus,

$$\bar{r}_n^* \approx 10^{-43}\,m.$$

From the Eq.(22) we can then conclude that the *primordial neutrons cluster* had a spherical shell with neutrons of *positive* gravitational masses and a spherical nucleus (inside the shell) with neutrons of *negative* gravitational masses (see Fig. 3).

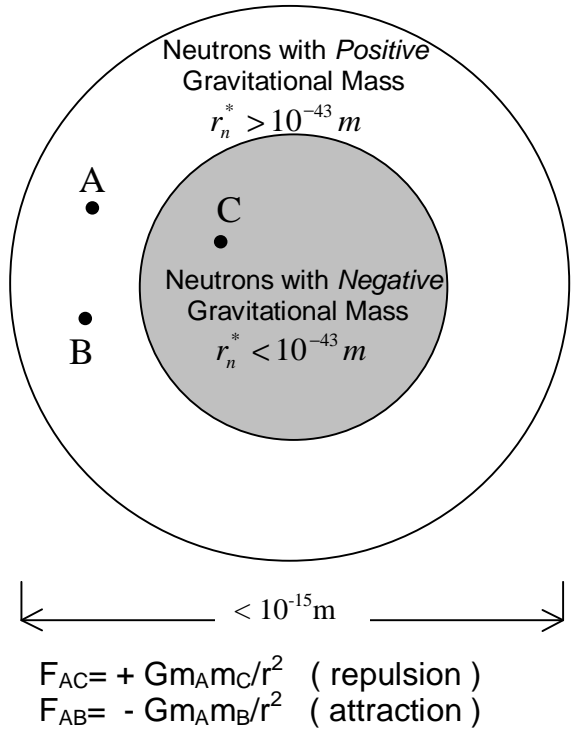

$F_{AC} = + Gm_Am_C/r^2$ ( repulsion )
$F_{AB} = - Gm_Am_B/r^2$ ( attraction )

Fig.3 - The *primordial neutrons cluster*

The negative gravitational masses neutrons' amount increase with the progress of the compression of the cluster. This means that the *gravitational repulsion* between the nucleus and the spherical shell also increase continuously with the compression of the cluster. Therefore there is a *critical point* in which the amount of negative gravitational mass is enough to explode the cluster. This is the experiment that we call the Big Bang.

This answers to the old cosmological question: If the universe is expanding, what was the origination of this expansion?

After the Big Bang, the neutrons' decompression began. The gravitational masses of the neutrons of the cluster's nucleus only become *positive* when their radii become greater than $6.3 \times 10^{-44} m$ (see Eq.22). The most compressed neutrons inside the cluster are in the central region of the nucleus, their supercompressing radii, $r_n^*$, are given by

$$r_n^* \approx r_0^* \cong r_{sn} = 2Gm_n/c^2 = 2.475 \times 10^{-54} m$$

When the radii of these neutrons increase to $6.3 \times 10^{-44} m$ the length scale of the Universe increases at the same proportion, i.e.,
$R/R_0 = 10^{-44}/10^{-54} \cong 10^{10}$
($R_0$ is the *initial* length scale of the Universe).On the other hand,
$R - R_0 = \frac{1}{2}\overline{g}t^2 \approx (GM_u/RR_0)t^2$
where $M_u \approx 10^{53} kg$ is the mass of the Universe.

This means that gravitation effectively remained repulsive for a period
$t = 10^{10}(R_0^3/GM_u)^{\frac{1}{2}} \approx 10^{-12} R_0^{\frac{3}{2}} << 10^{-34} s$,
after the Big Bang. During this period the Universe expanded at an astonishing rate, increasing its length scale by about a factor of $10^{10}$. Thus an isotropic *inflation* of our Universe ends at $\approx 10^{-36} s$.

Now a question: How did the *primordial neutrons* appear at the beginning of the Universe?

It is a proven *quantum* fact that a *wave function* $\Psi$ may *collapse* and that at this moment all the possibilities that it describes are suddenly expressed in *reality*. This means that, through this process, particles can be suddenly *materialized*.

The neutrons are created without its antiparticle ,the antineutron and thus it solves the matter/antimatter dilemma that is unresolved in many cosmologies .

The materialization of the primordial neutrons into a critical volume denotes *knowledge* of what would happen starting from that *initial condition*, fact that points towards the *existence* of a Creator.

Therefore, the Creator knew how the Universe would behave itself under *already existing laws.* Consequently, the laws were not created *for* the Universe and, hence, are not "Nature's laws" or "laws placed on Nature by the Creator", as written Descartes. They already existed as an intrinsic part of the Creator.

Thomas Aquinas had a very clear understanding about this. He talks about the Eternal Law "... which *exists* in God's mind and governs the whole Universe".

We then conclude that the Creator had all freedom to choose the initial conditions of the Universe, but opted for the concentration of the primordial neutrons into a critical volume so that the evolution of the Universe would proceed in the most convenient form for the purpose It had in mind and accordance with the laws inherent in Its own nature. This then answers the Einstein's famous question:" *What level of choice would God have had when building the Universe?*"

Apparently, Newton was the first to notice the Creator's option. In his book *Opticks*, he gives us a perfect view of how he imagined the creation of the Universe:

"*It seems possible to me that God, in the beginning, gave form to matter in solid, compacted particles(...) in the best manner possible to contribute to the purpose He had in mind ...*"

With what purpose did the Creator create the Universe? This question seems to be difficult to answer. Nevertheless, if we assume the Creator's primordial desire to *procreate*, i.e., to generate individual consciences from Itself so that the latter could envolve and manifest the same creating attributes pertaining to It, then we can infer that, in order for them to envolve, such consciences would need a Universe, and this might have been the main reason for its creation. Therefore, the origin of the Universe would be related to the generation of said consciences and, consequently, the materialization of the primordial neutrons must have taken place at the same epoch as when the Creator decided to *individualize* the postulated *primordial consciences*.

In this manner, not only is the *Material* Universe a creation of the Creator in Itself, but also the *Psychic* Universe. Thus, everything *is* in the Creator and It, being omnipresent, is in everything. Consequently, It cannot be displaced by another consciousness, not even by Itself, since It cannot, under the same aspect, be cause and effect at the same time, nor under an aspect, be cause of Itself and, under another, effect. Thus, the Creator is *immovable*.

As Augustine says (Gen. ad lit viii, 20), "The Creator Spirit moves Himself neither by time, nor by place."

Thomas Aquinas (Theology) also already had considered Creator's *immobility* as necessary.

For having been directly individualized into the Creator ( *Supreme Consciousness* ) the primordial consciousness certainly contained in themselves, although in a latent state, all the possibilities of the *Supreme Consciousness* , including the germ of independent will, which leads to creation of original starting points. However, in spite of the similarity to the *Supreme Consciousness*, the primordial consciousness could not have understanding of themselves. This understanding only arise with the creative mental state that consciences can only reach by evolution.

Thus, in the first evolutionary period, the primordial consciousness must have remained in total unconscious state, this being then the beginning of an evolutionary pilgrimage from *unconsciousness* to *superconscience*.